\documentclass[aps,graphics,showpacs]{revtex4}
\usepackage{graphicx}
\newtheorem{theorem}{Theorem}

\begin{document}
\title{Completely positive covariant two-qubit quantum processes and optimal quantum NOT operations for entangled
qubit pairs}

\author{J. Novotn\'y$^{(1)}$, G. Alber$^{(2)}$, I. Jex$^{(1)}$}
\affiliation{$^{(1)}$~Department of Physics, FJFI \v CVUT, B\v
rehov\'a 7,
115 19 Praha 1 - Star\'e M\v{e}sto, Czech Republic\\
$^{(2)}$~Institut f\"ur Angewandte Physik, Technische
Universit\"at Darmstadt, D-64289 Darmstadt, Germany}
\date{\today}

\begin{abstract}The structure of all completely positive quantum operations
is investigated which transform pure two-qubit input states of a given degree of entanglement
in a covariant way. Special cases thereof are quantum NOT operations which
transform entangled pure two-qubit input states of a given degree of entanglement into
orthogonal states in an optimal way. Based on our general analysis all covariant optimal
two-qubit quantum NOT operations are determined. In particular, it is demonstrated that only
in the case of maximally entangled input states these quantum NOT operations can be performed
perfectly.
\end{abstract}
 \maketitle

\section{Introduction}

The current interest in processing quantum information
\cite{Nielsen} has also revived the interest in fundamental
limitations of quantum theory \cite{Pati}. It is well known, for
example, that an arbitrary quantum state cannot be copied
perfectly \cite{nocloning}. Similar no-go theorems are also known
for other elementary tasks of quantum information processing
\cite{Pati}. One of these tasks, for example, concerns the problem
of transforming an arbitrary quantum state into an orthogonal one. It
is well accepted that for arbitrary (unknown) pure quantum states
such a quantum NOT operation cannot be performed perfectly due to
its anti-linear character \cite{Rungta,UNOT1}. In view of such
no-go theorems it is natural to investigate to which extent such
elementary tasks of quantum information processing can be
performed in an optimal way. In this context
quantum operations received
considerable
attention
which treat all possible
input states of interest in a covariant way \cite{UNOT1}. Such a
covariant behavior guarantees that the quantum process under
consideration achieves its goal for all input states of interest
with the same quality.

Recently, the problem of optimizing quantum NOT operations with
respect to arbitrary one-qubit input states stimulated both
theoretical \cite{Rungta} and experimental investigations
\cite{UNOT2}. By now many aspects of optimal quantum NOT
operations are well understood at least as far as general
one-qubit input states are concerned \cite{Rungta}. Nevertheless,
much less is known about optimal quantum NOT operations for
entangled input states. In particular, if one is interested in
constructing quantum NOT operations which are optimal for
entangled input states of a particular degree of entanglement
only, the general no-go theorem for quantum NOT operations does
not apply because the input states form a restricted subset and
not a linear subspace of the Hilbert space.

Motivated by these developments in this paper the problem of
constructing optimal quantum NOT operations for entangled quantum
states is addressed. In order to obtain a detailed first
understanding of this still open problem we concentrate our
discussion on the simplest possible input states, namely pure
two-qubit states of a given degree of entanglement. The main aim
of this paper is twofold. Firstly, the general structure of
completely positive quantum processes  is investigated
which transform all possible pure two-qubit inputs states of a
given degree of entanglement in a covariant way. Surprisingly it
turns out that all these processes can be represented in a
systematic way by  convex sums of four special quantum processes
some of which have already been discussed previously in the
literature. Secondly, based on this general analysis the structure
of two-qubit quantum processes is discussed which transform an
arbitrary pure two-qubit input state of a given degree of
entanglement into an orthogonal quantum state in an optimal way.
It is shown that in the special case of maximally entangled pure
input states  such quantum NOT operations can be performed
perfectly and the general structure of these perfect quantum NOT
operations is presented. These optimal quantum NOT operations may
have interesting future applications in the context of other
primitives of quantum information processing, such as remote state
preparation \cite{remote}. Finally, our work analyzes some of the
problems studied for a single qubit \cite{Ruskai} in the case of two qubits.

This paper is organized as follows: In Sec.II the most general structure of completely positive two-qubit
quantum processes is discussed which treat pure two-qubit input states of a given degree of entanglement
in a covariant way. The construction of optimal covariant quantum NOT operations and of perfect NOT operations
for maximally entangled input states are discussed in Sec.III. In Sec.IV the general
representation
of all possible completely positive covariant two-qubit processes is discussed once again. Thereby, it is demonstrated
that all these processes are convex sums of four special quantum operations whose physical significance is apparent
from the results obtained in Sec.III.

\section{Completely positive covariant two-qubit quantum processes\label{two-two}}

In this section the general structure of all completely positive
quantum processes is investigated which transform pure two-qubit input states of a
given degree of entanglement in a covariant way.

Let us start by considering a general
quantum operation \cite{Nielsen},
$\Pi$, which
maps an arbitrary two-qubit mixed input state,
$\rho_{in}$, onto a
mixed two-qubit output state, $\rho_{out}$, i.e.
\begin{equation}
\label{quantum_operation} \Pi: \hspace{1em} \rho_{in}
\longrightarrow \rho_{out}.
\end{equation}
If this is to treat pure two-qubit input states of a
given degree of entanglement in a covariant way it has to fulfill
the covariance condition \cite{covariance,Bruss}
\begin{equation}
\label{cavariant2} \Pi \left( U_1 \otimes U_2 \rho_{in}
U_1^{\dagger} \otimes U_2^{\dagger} \right) = U_1 \otimes U_2
\Pi(\rho_{in}) U_1^{\dagger} \otimes U_2^{\dagger}.
\end{equation}
This requirement has to be satisfied for arbitrary unitary
one-qubit transformations $U_1$,$U_2 \in SU(2)$ \cite{Biedenharn}.
The restriction of the quantum map (\ref{quantum_operation}) to
quantum operations reflects the physical requirement that $\Pi$
should be implementable by a unitary transformation possibly
involving also additional quantum systems but under the constraint
that initially the two-qubit system of interest and these
additional ancillary systems are uncorrelated \cite{Nielsen}. As
will be seen later, the covariance condition (\ref{cavariant2})
implies the requested independence of the quality of performance
of this quantum operation on the possible input states
\cite{Werner,Cerf}.

For implementing the covariance condition (\ref{cavariant2}) on
the quantum process of (\ref{quantum_operation}), it is convenient
to decompose the input state $\rho_{in}$ into its angular-momentum
irreducible tensor components $T(\frac{1}{2},\frac{1}{2})_{K,q}$
\cite{Blum}, i.e.
\begin{equation}
\label{rozklad}
\rho_{in}= \sum_{K,q;K^{'},q^{'}}
Tr\left\{\left[T\left(\frac{1}{2},\frac{1}{2}
\right)_{K,q}^{\dagger} \otimes
T\left(\frac{1}{2},\frac{1}{2}\right)_{K^{'},q^{'}}^{\dagger}\right]
\rho_{in}\right\}T\left(\frac{1}{2},\frac{1}{2}\right)_{K,q} \otimes
T\left(\frac{1}{2},\frac{1}{2}\right)_{K^{'},q^{'}}
\end{equation}
with
\begin{equation}
\label{tensor_components}
T\left(\frac{1}{2},\frac{1}{2}\right)_{0,0}=\frac{1}{\sqrt{2}}I, \hspace{1em}
T\left(\frac{1}{2},\frac{1}{2}\right)_{1,1}=-(\sigma_x+i\sigma_y)/2, \hspace{1em}
T\left(\frac{1}{2},\frac{1}{2}\right)_{1,0}=\sqrt{2}\sigma_z/2, \hspace{1em}
T\left(\frac{1}{2},\frac{1}{2}\right)_{1,-1}=(\sigma_x-i\sigma_y)/2
\end{equation}
and
with $K \in \{0,1\}$ and
$-K, -K+1,...\leq q \leq ...,K-1,K$.
Thereby,
$\sigma_i$ with $i=x,y,z$ are the three orthogonal components of the Pauli spin operators with
respect to fixed orthogonal $xyz$-axes.
(For the sake of convenience some basic facts about angular-momentum
tensor operators are summarized in Appendix \ref{tenzor}).
The corresponding most general linear covariant
output state has the form \cite{Werner,Nov1}
\begin{equation}
\label{general_output}
 \rho_{out}= \sum_{K,q;K^{'},q^{'}}
\lambda(K,K^{'})
Tr\left\{\left[T\left(\frac{1}{2},\frac{1}{2}\right)_{K,q}^{\dagger} \otimes
T\left(\frac{1}{2},\frac{1}{2}\right)_{K^{'},q^{'}}^{\dagger}\right]
\rho_{in}\right\}T\left(\frac{1}{2},\frac{1}{2}\right)_{K,q} \otimes
T\left(\frac{1}{2},\frac{1}{2}\right)_{K^{'},q^{'}}.
\end{equation}

According to equation (\ref{rozklad}) the most general two-qubit
input state can be written in the form
\begin{equation}
\label{vstup2} \rho_{in}(\vec{P},\vec{Q},\mathcal{M})= \frac{1}{4}
\left\{ I\otimes I + \sum_{i=x,y,z} P_i \sigma_i \otimes I +
\sum_{i=x,y,z} Q_i I \otimes \sigma_i + \sum_{ i,j=x,y,z} M_{ij}
\sigma_{i} \otimes \sigma_{j} \right\}.
\end{equation}
with the aid of the two local vectors of coherence,
$\vec{P}=(P_x,P_y,P_z)$ and $\vec{Q}=(Q_x,Q_y,Q_z)$,  and with the
correlation-tensor $\mathcal{M}=(M_{ij})_{i,j=x,y,z}$
\cite{Schlienz}. Because we are looking for trace preserving maps,
we obtain the condition $\lambda(0,0)=1$. Using the notation
$V=\lambda(1,0)$, $X=\lambda(0,1)$, $Y=\lambda(1,1)$ the
corresponding output state of (\ref{general_output}) is given by
\begin{equation}
\label{output2}
 \rho_{out} = \frac{1}{4} \left\{ I\otimes I +
\sum_{i=x,y,z} (VP_i) \sigma_i \otimes I + \sum_{i=x,y,z} (XQ_i) I
\otimes \sigma_i + \sum_{ i,j=x,y,z} (YM_{ij}) \sigma_{i} \otimes
\sigma_{j} \right\} \equiv \rho_{in}(V\vec{P},X\vec{Q},Y\mathcal{M}).
\end{equation}
In the special case of
a normalized pure input state $|\psi\rangle =
\alpha |\uparrow \uparrow \rangle + \beta |\downarrow \downarrow
\rangle$ which is quantized in the $z$-direction this yields the explicit matrix representation
\begin{equation}
\label{pure_output2} \rho_{out}=\left( \begin{array}{cccc}
\frac{1+Y}{4}+\frac{X+V}{4}(|\alpha|^2-|\beta|^2)&0&0&Y\alpha\beta^{*} \\
0&\frac{1-Y}{4}+\frac{V-X}{4}(|\alpha|^2-|\beta|^2)&0&0 \\
0&0&\frac{1-Y}{4}+\frac{X-V}{4}(|\alpha|^2-|\beta|^2)&0 \\
Y\alpha^{*}\beta&0&0&\frac{1+Y}{4}-\frac{X+V}{4}(|\alpha|^2-|\beta|^2)
\end{array} \right)
\end{equation}
in the eigenbasis of $\sigma_z\otimes \sigma_z$.
Therefore,
an arbitrary triple $(X,V,Y)$
defines the most general covariant
map between an input state (\ref{vstup2}) and an output state (\ref{output2}).
Further restrictions are imposed on these parameters by complete positivity \cite{Ruskai}.
As shown in detail in Appendix \ref{Appendix_a2}, complete positivity requires that
all components of the triple $(X,V,Y)$ have to be real-valued and they have to fulfill
the relations
\begin{equation}
\label{relace_positivity}
 1+3X+3V+9Y \geq 0, \hspace{1em}
1+3X-V-3Y \geq 0, \hspace{1em} 1-X+3V-3Y \geq 0, \hspace{1em}
1-X-V+Y \geq 0,
\end{equation}
or equivalently
\begin{equation}
-\frac{1}{3} \leq X,V \leq 1, \hspace{2em} \max\left\{
-\frac{1+3X+3V}{9},-1+X+V\right\} \leq Y \leq \frac{1+3\min\{X,V\}
- \max\{X,V\}}{3}.
\end{equation}
Thus,
provided
these relations are fulfilled the
process defined by the covariant output state (\ref{output2}) is
completely positive. A Kraus-representation of this
deterministic quantum operation is given by
\begin{equation}
\label{krausform} \rho_{out}=
\Pi_{V,X,Y}\left(\rho_{in}(\vec{P},\vec{Q},\mathcal{M})\right)=
\sum_{i,j=0,x,y,z} K_{ij} \rho_{in}(\vec{P},\vec{Q},\mathcal{M})
K_{ij}^{\dagger} = \rho_{in}(V\vec{P},X\vec{Q},Y\mathcal{M})
\end{equation}
with
\begin{eqnarray}
K_{00}&=&\frac{1}{4}\left(1+3X+3V+9Y\right)^{\frac{1}{2}} I
\otimes I, \hspace{1em} K_{i0}=
\frac{1}{4}\left(1+3X-V-3Y\right)^{\frac{1}{2}} \sigma_i \otimes I, \nonumber \\
K_{0i}&=&\frac{1}{4}\left(1-X+3V-3Y\right)^{\frac{1}{2}} I \otimes
\sigma_i, \hspace{1em}
K_{ij}=\frac{1}{4}\left(1-X-V+Y\right)^{\frac{1}{2}} \sigma_i
\otimes \sigma_j, \hspace{1em} i,j \in \{ x,y,z \}.
\end{eqnarray}
Trace preservation is implied by the relation
\begin{equation}
\sum_{i,j=0,x,y,z} K_{ij}^{\dagger}K_{ij}=I.
\end{equation}

\begin{figure}
\includegraphics[width=12.cm]{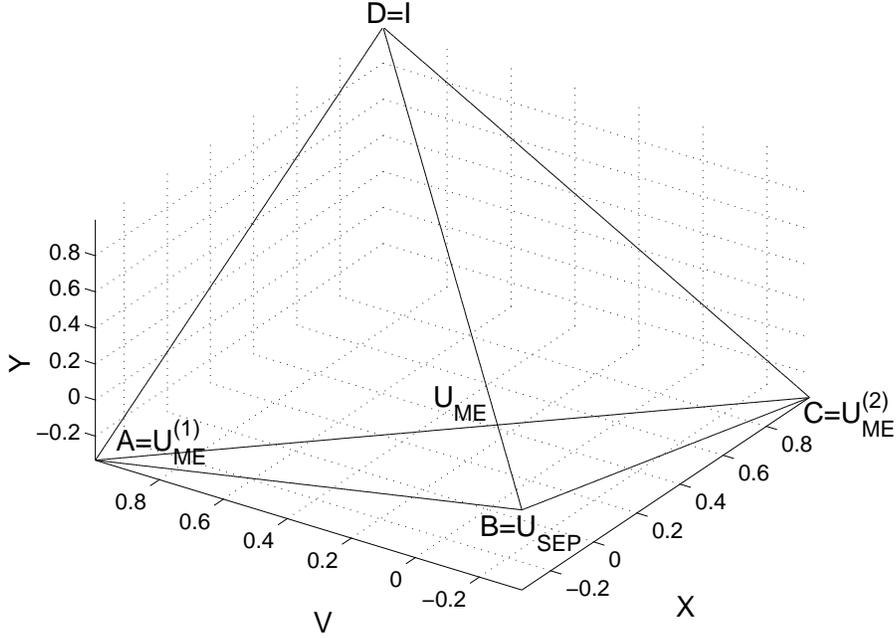} \caption{The parameter space of points
$(V,X,Y)$ for which the covariant process $\Pi_{V,X,Y}$ is
completely positive forms the tetrahedron $\overline{ABCD}$.}
\label{positivita}
\end{figure}

The set of all possible completely positive universal quantum operations
characterized by triples $(V,X,Y)$
is represented by the convex tetrahedron $\overline{ABCD}$
of Fig.
\ref{positivita}.
The physical significance of the
extremal points of this tetrahedron is discussed in Sec.\ref{representation}.

\section{Optimal quantum NOT operations for pure entangled qubit pairs\label{Optimal}}

Starting from the general results of Sec.\ref{two-two} we can
specify different types of completely positive covariant quantum
processes. In the following we determine quantum processes which
describe a quantum NOT operation acting on arbitrary pure
two-qubit states of a given degree of entanglement in an optimal
way.

Let us first of all summarize the basic problems which arise if one wants
to construct a quantum NOT operation for arbitrary input states of a
complex Hilbert space $\mathcal{H}$. Such a quantum NOT operation has to
map an
arbitrary pure input state $|\phi \rangle \in \mathcal{H}$ onto another
pure
orthogonal state $|\phi
\rangle^{\bot} \in \mathcal{H}$ in such a way that $\langle \phi|\phi^{\bot}\rangle=0$
holds. An ideal quantum NOT operation has to be anti-linear \cite{UNOT1}  and hence
it is not possible to represent its operation by a complete positive quantum
operation.
In view of this no-go property of quantum mechanics it is of interest to construct
quantum operations which approximate a quantum NOT operations in the best possible way
only for a restricted class of input states.

One of the simplest examples in this context
is the construction of an optimal quantum NOT operation for pure two-qubit states of a given degree of
entanglement. For this purpose it is convenient first of all to
decompose the relevant four dimensional
Hilbert space $\mathcal{H}$ of two qubits into the possible classes
of pure two-qubit states
\begin{equation}
\Omega_{\alpha}= \Big\{ \big(U_{1} \otimes U_{2}\big) \big( \alpha
|\uparrow\rangle \otimes  |\uparrow \ \rangle + \beta
|\downarrow\rangle \otimes |\downarrow \rangle \big) \Big| U_{1},
U_{2} \in {SU}(2)  \Big\} \label{Omega}
\end{equation}
with the same degree of entanglement. Thereby, the parameter
$\alpha$ ($0 \leq \alpha \leq \frac{1}{\sqrt{2}}, \beta = \sqrt{1-
\alpha^2}$) characterizes the degree of entanglement of the pure
states in a given class $\Omega_{\alpha}$ \cite{Nov1,Bruss}. Note
that in the special case $\alpha =0$ the two-qubit state is
separable (SEP) whereas in the opposite extreme case $\alpha =
1/\sqrt{2}$ it is maximally entangled (ME). We are interested in
constructing linear and completely positive quantum processes
$U_{\alpha}$ which map an arbitrary pure input state, say $|\phi
\rangle \in \Omega_{\alpha}$, in an optimal way onto its
orthogonal complement, i.e.
\begin{equation}
U_{\alpha} : \rho_{in} = |\phi \rangle \langle \phi |
\longrightarrow \rho_{out}. \label{map}
\end{equation}

For the solution of this optimization problem a measure is needed
which quantifies how close the output state $\rho_{out}$ is to the
orthogonal complement of the input state $|\phi\rangle$.
Definitely, the Hilbert space of two qubits $\mathcal{H}$ is the
direct sum of two Hilbert spaces, namely the span of the vector
$|\phi\rangle$, say $\mathcal{H_{\phi}}$, and its
three-dimensional orthogonal complement
$\mathcal{H_{\phi}}^{\bot}$. Therefore, a convenient measure is
given by the minimal distance between the output state
$\rho_{out}$ and all mixed states contained in the orthogonal
complement of the input, i.e.
\begin{equation}
\label{distance1} D(\rho_{out}|\phi^{\bot})=\min_{\sigma \in
\Gamma(\mathcal{H_{\phi}}^{\bot})} Tr\{\rho_{out} - \sigma\}^2.
\end{equation}
Thereby, $\Gamma(\mathcal{H^{\bot}_{\phi}})$ denotes the linear
convex set of all density operators formed by convex sums of pure
states of the Hilbert space $\mathcal{H_{\phi}^{\bot}}$. This
measure is based on the well known Hilbert-Schmidt norm for
Hilbert-Schmidt operators $A$ and $B$, i.e. $||A - B|| =
\sqrt{Tr\left\{A-B\right\}^2}$. We omitted the square root as it
is unimportant for our purposes. As shown in Appendix
\ref{appendix2a}, the minimal distance of (\ref{distance1}) can
also be express in the more convenient form
\begin{equation}
\label{distance2}
D(\rho_{out}|\phi^{\bot})=2\langle\phi|\rho_{out}^2|\phi\rangle
-\frac{2}{3} \langle\phi|\rho_{out}|\phi\rangle^2.
\end{equation}
Correspondingly,
the largest achievable distance, i.e.
\begin{equation}
\label{fid} \Delta(U_{\alpha})= \sup_{\phi \in
\Omega_{\alpha}}D(\rho_{out}|\phi^{\bot})=\sup_{\phi \in
\Omega_{\alpha}} \left\{2\langle\phi|\rho_{out}^2|\phi\rangle
-\frac{2}{3} \langle\phi|\rho_{out}|\phi\rangle^2 \right\},
\end{equation}
is a convenient error measure characterizing the quality of the
NOT operation for a given class of input states with a given
degree of entanglement. This error measure has two important
properties (for details see Appendix \ref{appendix2b}). Firstly,
the positivity of density operators implies that it is zero if and
only if the NOT operation is ideal for all input states
$|\phi\rangle \in \Omega_{\alpha}$, i.e.
\begin{equation}
\Delta(U_{\alpha})=0 \hspace{2em} \iff \sup_{| \phi \rangle \in
\Omega_{\alpha}} \langle \phi | \rho_{out} | \phi \rangle =0.
\end{equation}
Secondly, this error measure is invariant under the unitary group
$U(4)$. For the covariant processes of (\ref{krausform}) this
implies that the distance $D(\rho_{out},\phi^{\bot})$ is unbiased
with respect to all states from the class $\Omega_{\alpha}$. Thus,
for these processes we can omit the supremum in (\ref{fid}) and we
can calculate the error as the distance (\ref{distance2})
associated with an arbitrarily chosen state of the class
$\Omega_{\alpha}$.

Therefore, in general the construction of an optimal quantum NOT
operation is equivalent to minimizing the error measure
$\Delta(U_{\alpha})$ over all possible processes. In the following
the resulting optimal error measure will be denoted by
$\Delta_{\alpha} = \inf_{U_{\alpha}} \Delta(U_{\alpha})$.

\subsubsection{Non-covariant quantum NOT operations for maximally entangled qubit pairs}

Before dealing with the general case let us focus on quantum NOT
operations for the special class of maximally entangled (ME) pure
input states $\Omega_{\-1/\sqrt{2}}$. In this special case one is
able to construct even perfect quantum NOT operations which map an
arbitrary pure input state onto a pure output state but which are
typically not covariant.

In order to determine the general structure of all physically
feasible quantum NOT operations $\mathcal{U}$ for ME states let us
impose the natural additional requirement that, if the quantum NOT
operation $\mathcal{U}$ is applied twice the resulting operation
is proportional to the identity operator. Therefore, the quantum
NOT operation $\mathcal{U}$ we are looking for should fulfill the
following requirements:
\begin{itemize}
\item{Orthogonality: It maps an arbitrary pure state onto
a pure state according to}
\begin{equation}
\label{NOT} \langle\phi |\mathcal{U}|\phi \rangle =0
\hspace{2em}\hspace{2em} \forall
\hspace{0.3em}|\phi\rangle \in \Omega_{1/\sqrt{2}}
\end{equation}
\item{Unitarity}
\begin{equation}
\label{unitary} \mathcal{U}\mathcal{U}^{\dagger} = I
\end{equation}
\item{Cyclic property}
\begin{equation}
\label{double} \mathcal{U}^{2}=\lambda I, \hspace{2em} {\rm where}
\hspace{1em} \lambda \in {\mathbf C}.
\end{equation}
\end{itemize}

For our analysis we take advantage of the special basis states
(sometimes referred to as the magic base) \cite{Cerf}
\begin{eqnarray}
|e_{1}\rangle = \frac{1}{\sqrt{2}} \left( |00\rangle +
|11\rangle\right), \hspace{1em} |e_2\rangle =
\frac{i}{\sqrt{2}}\left(|00\rangle - |11\rangle\right),
\hspace{1em} |e_3\rangle = \frac{i}{\sqrt{2}}\left(|01\rangle +
|10\rangle\right), \hspace{1em} |e_4\rangle =
\frac{1}{\sqrt{2}}\left(|01\rangle-|10\rangle\right),
\end{eqnarray}
in which all maximally entangled two-qubit states can be written
as real-valued linear combination of these basis states. Clearly,
the concurrence of an arbitrary normalized two-qubit superposition
state $|\Gamma\rangle = \sum_{i}\gamma_i |e_i\rangle$ with complex
values of $\gamma_i$ is given by
\begin{equation}
C\left(|\Gamma\rangle \langle \Gamma |\right) = \left|\sum_i
\gamma_i^2\right|.
\end{equation}
Hence, for ME states this concurrence has to be equal to unity. This happens
if and only if all coefficients $\gamma_i$ are real-valued. In this sense all
ME states form a four dimensional real Hilbert space.
Expressing condition (\ref{NOT}) in this magic base it turns out
that all possible quantum NOT operations form a vector space of real-valued
4x4 antisymmetric matrices. The dimension of this
vector space equals six  and a possible basis is given by the matrices
\begin{eqnarray}
U_1&=&\left( \begin{array}{cccc}
0&1&0&0\\-1&0&0&0\\0&0&0&1\\0&0&-1&0
\end{array}\right), \hspace{2em} U_2=\left( \begin{array}{cccc}
0&0&0&1\\0&0&1&0\\0&-1&0&0\\-1&0&0&0
\end{array}\right), \hspace{2em} U_3=\left( \begin{array}{cccc}
0&0&1&0\\0&0&0&-1\\-1&0&0&0\\0&1&0&0
\end{array}\right), \\
V_1&=&\left( \begin{array}{cccc}
0&1&0&0\\-1&0&0&0\\0&0&0&-1\\0&0&1&0
\end{array}\right), \hspace{2em} V_2=\left( \begin{array}{cccc}
0&0&0&1\\0&0&-1&0\\0&1&0&0\\-1&0&0&0
\end{array}\right), \hspace{2em} V_3=\left( \begin{array}{cccc}
0&0&1&0\\0&0&0&1\\-1&0&0&0\\0&-1&0&0
\end{array}\right).
\end{eqnarray}
This set of matrices has the following interesting algebraic
properties
\begin{eqnarray}
\{U_i,U_j^{\dagger}\} &=& -\left\{U_i,U_j\right\} = 2\delta_{ij}I,
\hspace{2em} U_i^{T}=-U_i, \hspace{2em} U_iU_j=-\delta_{ij}I +
\varepsilon_{ijk}U_k, \nonumber \\
\{V_i,V_j^{\dagger}\} &=& -\left\{V_i,V_j\right\}= 2\delta_{ij}I,
\hspace{2em} V_i^{T}=-V_i, \hspace{2em} V_iV_j=-\delta_{ij}I +
\varepsilon_{ijk}V_k, \hspace{2em} [U_{i},V_{j}]=0. 
\label{property}
\end{eqnarray}
As a consequence
every linear operation with the property (\ref{NOT}) is a linear
superposition of $U_i$, $V_i$, i.e.
\begin{equation}
\mathcal{U}= \sum_{i=1}^{3} \alpha_i U_i + \beta_i V_i, \hspace
{2em} \alpha_i,\beta_i \in {\mathbf R}.
\end{equation}
Property (\ref{property}) and requirement (\ref{unitary})
imply the relation
\begin{eqnarray}
I &=& {\mathcal U} {\mathcal U}^{\dagger} = \sum_{i=1}^{3}
\alpha_i^2U_iU_i^{\dagger} + \beta_i^2 V_iV_i^{\dagger} +
\sum_{(ij)} \left[\alpha_i\alpha_j ( U_iU_j^{\dagger} +
U_j^{\dagger}U_i)+\beta_i\beta_j( V_iV_j^{\dagger} +
V_j^{\dagger}V_i )\right]+ \sum_{i,j =1}^3
\alpha_i \beta_j (U_iV_j^{\dagger} + V_jU_i^{\dagger}) \nonumber\\
&=& \left(\sum_{i=1}^3 \alpha_i^2 + \beta_i^2\right) I -
2\sum_{i,j=1}^3 \alpha_i \beta_j U_iV_j.
\end{eqnarray}
Taking into account the structure of the matrices
$U_iV_j$ this yields the conditions
\begin{equation}
\alpha_i \beta_j=0 \hspace{2em} \quad \Longrightarrow
\left\{\begin{array}{c}\alpha_i=0
\quad \wedge \quad \sum\limits_{i=1}^3 \beta_i^2 = 1\\
\beta_i=0 
\quad \wedge \quad \sum\limits_{i=1}^3 \alpha_i^2 = 1
\end{array}\right..
\end{equation}
The quantum NOT operation fulfilling requirements (\ref{NOT}),
(\ref{unitary}) and (\ref{double}) has the general structure
\begin{equation}
\label{ideal_NOT} \left( \mathcal{U}= \sum_{i=1}^3 \alpha_i U_i,
\quad {\rm with} \quad \sum\limits_{i=1}^3 \alpha_i^2=1,\alpha_i\in
{\mathbf R}\right) \qquad \vee \qquad \left( \mathcal{U}=
\sum\limits_{i=1}^3 \beta_i V_i, \quad {\rm with} \quad
\sum\limits_{i=1}^3 \beta_i^2=1,\beta_i\in {\mathbf R}\right).
\end{equation}
In both cases
the condition  (\ref{double})is fulfilled automatically, i.e.
\begin{equation}
\mathcal{U}^2= -I.
\end{equation}
Therefore, for maximally entangled two-qubit states the ideal
quantum NOT operation is not unique. Its most general form is given
by (\ref{ideal_NOT}).

\subsubsection{Optimal covariant quantum NOT operations}

Let us construct optimal quantum NOT operations for arbitrary
classes of pure two-qubit input states of a given degree of
entanglement $\Omega_{\alpha}$. In this case a similar strategy
can be used as the one used for the construction of optimal
universal quantum copying processes \cite{Werner}. Similarly, it
can be shown (for details see Appendix \ref{appendix2b}) that for
any optimal quantum NOT operation $U_{\alpha}$ always an
equivalent covariant quantum process (\ref{krausform}), say
$\widehat{U_{\alpha}}$, can be found which fulfills the covariance
condition (\ref{cavariant2}). Thus, this latter quantum NOT
process yields the same optimal error measure $\Delta_{\alpha}$
for all possible two-qubit input states $|\phi\rangle \in
\Omega_{\alpha}$. This basic observation allows us to restrict our
search for the optimal quantum NOT operation for an arbitrary
class $\Omega_{\alpha}$ to covariant quantum processes of the form
of (\ref{krausform}) which minimize the error measure (\ref{fid}).

The error measure of the output state
(\ref{pure_output2}) with respect to the normalized pure two-qubit
input state
$|\phi\rangle={\alpha|\uparrow\uparrow\rangle} +
{\beta|\downarrow\downarrow\rangle}$ is given by
\begin{equation}
\label{fidelity_unot} \Delta(Z=V+X,Y)= \frac{1}{12}\left\{
\left[1+Z(1-4\alpha^2\beta^2)+Y(1+8\alpha^2\beta^2)\right]^2
+6\alpha^2\beta^2(1-4\alpha^2\beta^2)(Z-2Y)^2\right\}.
\end{equation}
In Appendix \ref{appendix3} it is shown that
for all classes of states $\Omega_{\alpha}$
all optimal quantum NOT processes are
determined by points $(V,X,Y)$ of the
triangle $\overline{ABC}$ of Fig.\ref{positivita}. Therefore,
for an optimal quantum NOT process the operator $K_{00}$
of the Kraus representation (\ref{krausform}) vanishes.
Thus,
minimizing the quantity (\ref{fidelity_unot}) with
respect to points of the triangle $\overline{ABC}$ yields the final solution.
Depending on the value of $\alpha$
two cases can be distinguished.
For $\alpha \leq \alpha_0$  with
$\alpha_0=\sqrt{\frac{1-\sqrt{1-4K}}{2}}\approx 0.1836$ and
$K=\frac{8-3\sqrt{6}}{20}$
the minimal error
\begin{equation}
\Delta_{\alpha}=\frac{1}{243}
\left(4+160\alpha^2\beta^2-128\alpha^4\beta^4\right)
\end{equation}
is obtained.
The resulting optimal quantum NOT operation
is independent of the parameter $\alpha$ and is characterized by the
point
$(V=-\frac{1}{3},X=-\frac{1}{3},Y=\frac{1}{9})$. It turns out that
this particular optimal
quantum NOT process $U_{SEP}$ consists of two
one-qubit optimal covariant U-NOT processes $u^1$ applied to each of the qubits
separately, i.e.
$U_{SEP}=u^1 \otimes u^1$ with
\begin{equation}
u^1(\rho)=\frac{1}{3}\left( 2I-\rho \right).
\end{equation}
These latter optimal one-qubit U-NOT quantum processes were
studied in detail in \cite{UNOT1}. According to (\ref{krausform})
a Kraus representation of the optimal two-qubit quantum NOT
operation $U_{SEP}$ is given by
\begin{equation}
U_{SEP}(\rho_{in}) = \sum_{i,j=1}^{3} K_{ij}
\rho_{in}K_{ij}^{\dagger} \qquad {\rm with} \quad K_{ij} =
\frac{1}{3} \sigma_i \otimes \sigma_j.
\end{equation}
Optimal quantum NOT processes with
$\alpha \geq \alpha_{0}$ yield an error of magnitude
\begin{equation}
\label{error_notsep}
\Delta_{\alpha}=\frac{4\alpha^2\beta^2(1-4\alpha^2\beta^2)}{2+35\alpha^2\beta^2-100\alpha^4\beta^4}
\end{equation}
and they are characterized by points
$(V,X,Y)$ on the straight line
\begin{equation}
\label{parametry}
Y=-\frac{1}{3}\frac{2-31\alpha^2\beta^2+20\alpha^4\beta^4}{-2-35\alpha^2\beta^2+100\alpha^4\beta^4},
\hspace{2em}
X+V=Z=\frac{2}{3}\frac{4-29\alpha^2\beta^2-20\alpha^4\beta^4}{-2-35\alpha^2\beta^2+100\alpha^4\beta^4},
\hspace{2em} X,V \geq -\frac{1}{3}.
\end{equation}
Each triple of parameters $(V,X,Y)$ from this one-parameter line
segment defines the Kraus representation (\ref{krausform}) of the
optimal two-qubit quantum NOT operation $\widehat{U_{\alpha}}(V)$
for a particular class of states $\Omega_{\alpha}$.

These considerations show that an ideal covariant two-qubit
quantum NOT process with  zero-valued error measure can only be
obtained for maximally entangled states. Such a process is
characterized by any point $(V,X,Y)$ satisfying the conditions
$Y=-\frac{1}{3}, X+V=Z=\frac{2}{3}$, ($X,V \geq -\frac{1}{3}$).
Therefore, ideal covariant two-qubit quantum NOT processes form a
one-parameter family. This reflects the fact that there is a
huge class of non-covariant ideal quantum NOT operations
(\ref{ideal_NOT}). Each element $\mathcal{U}$ of this class
corresponds to some covariant counterpart $\widehat{\mathcal{U}}$
with the same error (\ref{fid}) (see Appendix \ref{appendix2b}).
Thus, for maximally entangled states the ideal covariant two-qubit
NOT operations are characterized by the parameter range $-\frac{1}{3}
\leq V \leq 1$. A Kraus representation of these processes is given
by
\begin{equation}
U_{ME}(V) (\rho_{in}) = \sum_{i=1}^{3} \left(K_{0i} \rho_{in}
K_{0i}^{\dagger} + K_{i0}\rho_{in}K_{i0}^{\dagger} \right),
\end{equation}
with
\begin{equation}
K_{0i}=\frac{1}{2} \left(\frac{1}{3}+V\right)^{1/2} \sigma_i
\otimes I,  \quad K_{i0}=\frac{1}{2} \left(1-V\right)^{1/2} I
\otimes \sigma_i.
\end{equation}

The error $\Delta_{\alpha}$ achieves its maximal value for
$\alpha^2\beta^2=\frac{1}{10}$, i.e.
$\alpha_{max}=\sqrt{\frac{1}{2}-\sqrt{\frac{3}{20}}}$. The
corresponding maximal error is given by
$\Delta_{\alpha_{max}}=\frac{4}{75}$ and its associated quantum
processes are characterized by the points $(V,X,Y)$ with
$Y=-\frac{1}{15}$ and $X+V=Z=-\frac{2}{15}$ ($X,V \geq
-\frac{1}{3}$). One of the processes satisfying these conditions
is the four-dimensional covariant U-NOT process
$\mathcal{G}_{NOT}$ introduced in Ref.\cite{Rungta}. This particular
covariant two-qubit U-NOT process minimizes the error with respect
to all possible two-qubit pure input states independent of their
degree of entanglement. This special process is characterized by
the parameters $X=V=Y=-\frac{1}{15}$ and it maps an arbitrary
two-qubit input state $\rho$ onto the output state
\begin{equation}
\rho_{out} = \mathcal{G}_{NOT}(\rho) = \frac{1}{15}\left( 4I -
\rho \right).
\end{equation}

\begin{figure}
\includegraphics[width=11.cm]{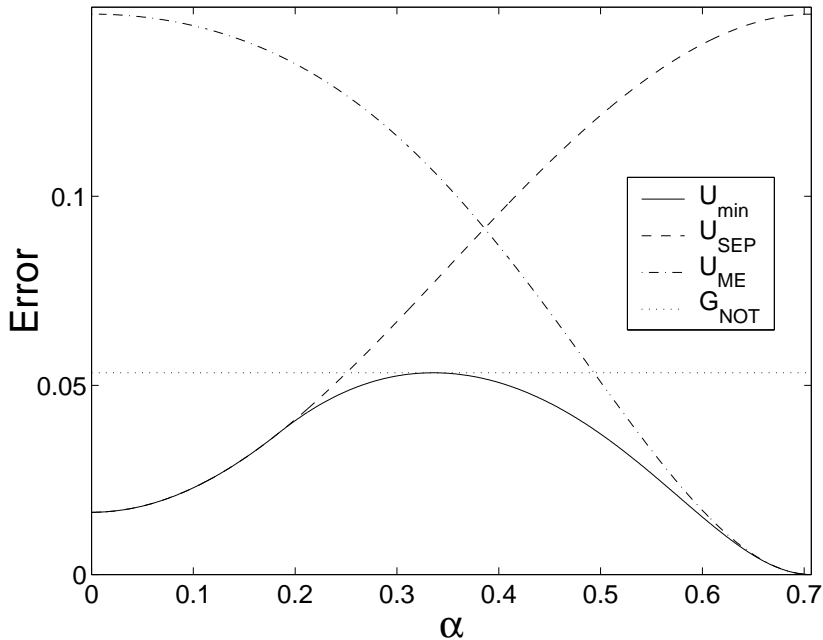}
\caption{The minimum error (\ref{fid}) and the errors of the three
relevant U-NOT processes and their dependence on the degree of
entanglement $\alpha$. The solid line represents the optimal
minimum error. The dashed line $U_{SEP}$ corresponds to an
independent application of two one-qubit covariant U-NOT
operations $u^1$ to each qubit from the entangled pair. The dashed-dotted
line $U_{ME}$ corresponds to the ideal covariant U-NOT
map for maximally entangled states. The dotted line represents the
minimum achievable error for an unknown two-qubit pure state if
its degree of entanglement is unknown.} \label{fidelity3}
\end{figure}

In summary,
the smallest achievable errors $\Delta_{\alpha}$ for these
optimal covariant two-qubit quantum NOT processes $\hat{U_{\alpha}}$ are given by
\begin{equation}
\Delta_{\alpha} = \left\{ \begin{array}{ccc} \frac{1}{243}
\left(4+160\alpha^2\beta^2-128\alpha^4\beta^4\right), &
U_{SEP}=u^1
\otimes u^1, & {\rm for} \quad \alpha \leq \alpha_0 \nonumber \\
\frac{4\alpha^2\beta^2(1-4\alpha^2\beta^2)}{2+35\alpha^2\beta^2-100\alpha^4\beta^4},
& \widehat{U_{\alpha}}(V), & {\rm for} \quad \alpha \geq \alpha_0
\nonumber \\
\frac{4}{75}, & U_{\alpha_{max}}=\mathcal{G}_{NOT}, & {\rm for}
\quad \alpha=\alpha_{max} \nonumber \\
0, & U_{ME}(V), & {\rm for} \quad \alpha=\frac{1}{\sqrt{2}}
\end{array} \right.
\end{equation}
Their dependence on the degree of entanglement $\alpha$ is depicted in Fig.
{\ref{fidelity3}}.

The optimal way to complement two-qubit pure separable states with
$\alpha=0$ is to perform one-qubit covariant U-NOT quantum
operations on each qubit independently. The resulting minimum
error for separable states is given by $\Delta_0=\frac{4}{243}$.
This quantum process also yields the minimal error for two-qubit
pure states with $\alpha \leq \alpha_0$. But the minimum error
$\Delta_{\alpha}$ increases monotonically with the degree of
entanglement up to the critical value $\alpha_0 \approx 0.1836$
with $\Delta_{\alpha_0} \approx 0.0373$. For $\alpha \geq
\alpha_0$ the covariant processes $\widehat{U_{\alpha}}(V)$ are
optimal. These processes reach their maximum error at
$\alpha_{max}= \sqrt{\frac{1}{2}-\sqrt{\frac{3}{20}}}$ and for
maximally entangled states with  $\alpha = 1/\sqrt{2}$ the error
vanishes.

These results demonstrate
that only in the case of ME states one is able to construct ideal
covariant quantum NOT
processes. This implies that there are no non-covariant ideal
quantum
NOT processes for non-maximally entangled pure
states. This can be proved indirectly.
Suppose that such processes existed. In this case we were able to construct
to each ideal non-covariant
quantum NOT
process a corresponding covariant process
(compare with Appendix \ref{appendix2b}). However, this is in
direct contradiction with our findings.
Moreover, this fact also tells us that there is no magic
base for sets of states $\Omega_{\alpha}$ ($\alpha \neq
\frac{1}{\sqrt{2}}$). Only maximally entangled states make up a
real subspace of the Hilbert space of two qubits. This emphasizes once more the
special character of the set of maximally entangled states
in comparison with all other pure entangled states.

\section{General representation of universal two-qubit processes \label{representation}}

Based on the results of Sec.\ref{Optimal} all possible completely
positive covariant two-qubit processes as defined by
(\ref{krausform}) can be represented by convex combinations of four
basic quantum processes which correspond to the corners of the
tetrahedron $\overline{ABCD}$ of Fig.\ref{positivita}. For this
purpose let us briefly summarize the graphical representation of
these completely positive covariant quantum maps. According to the
results of Appendix \ref{appendix3} all optimal two-qubit quantum
NOT operations have to be presented by points of the triangle
$\overline{ABC}$. Thereby, point
$B=(V=-\frac{1}{3},X=-\frac{1}{3},Y=\frac{1}{9})$ characterizes a
quantum NOT operation minimizing the error (\ref{fid}) for classes
of states $\Omega_{\alpha}$ with $\alpha \leq \alpha_{0}$.  Points
on straight lines specified by the parameters (\ref{parametry})
characterize optimal quantum NOT processes minimizing the error
(\ref{fid}) for the classes of states $\Omega_{\alpha}$ with $\alpha
\geq \alpha_{0}$. In particular, points with
$Y=-\frac{1}{3},X+V=Z=\frac{2}{3}$,($X,V \geq -\frac{1}{3}$) define
optimal quantum NOT processes for maximally entangled states. The
line segments $\overline{AD}$ and $\overline{CD}$ correspond to the
restrictions $V=1$ and $X=1$. Therefore, they specify completely
positive covariant processes which do not change the reduced density
operator of the first or the second qubit. The process corresponding
to the point $D$ leaves both reduced density operators unchanged.
So, it represents the identity operations. Furthermore, the
processes represented by the points $(A = V=1, X=-\frac{1}{3},
Y=-\frac{1}{3})$ and $(C=V=-\frac{1}{3}, X=1, Y=-\frac{1}{3})$ are
ideal covariant quantum NOT operations for maximally entangled
states and moreover they do not change the reduced density operators
of the first and second qubit. Therefore, we have the
correspondences
\begin{equation}
\label{convex_base} U_{ME}^{(1)}~~\longleftrightarrow~~A, \hspace{2em}
U_{SEP}~~\longleftrightarrow~~B, \hspace{2em}
U_{ME}^{(2)}~~\longleftrightarrow~~C, \hspace{2em} I~~\longleftrightarrow~~D.
\end{equation}
In terms of these special quantum processes
all possible completely positive covariant two-qubit processes can be represented as convex combinations.
Thus, a two-qubit quantum
operation (\ref{quantum_operation}) is completely positive and
fulfills the covariance condition (\ref{cavariant2}) if and only
if it can be expressed as a linear convex combination of these basic quantum operations, i.e.
(\ref{convex_base})
\begin{equation}
\Pi_{a_1,a_2,a_3,a_4} = a_1I + a_2U_{SEP} + a_3U_{ME}^{(1)} +
a_4U_{ME}^{(2)}, \hspace{2em} a_i \geq 0 \hspace{1em} {\rm and}
\hspace{1em} \sum_{i=1}^4 a_i =1.
\end{equation}

\section{Conclusion}

A classification of all possible completely positive covariant
two-qubit quantum processes was presented which fulfill the
covariance condition (\ref{cavariant2}). It could be shown that any of
these processes can be represented by a convex sum of four special
covariant two-qubit quantum processes some of which had already
been discussed in the literature previously. On the basis of this general
classification all possible completely positive covariant quantum
processes were constructed which describe quantum NOT operations
acting on pure two-qubit states of a particular degree of
entanglement in an optimal way. It was shown that for maximally
entangled pure two-qubit input states even an ideal covariant
quantum NOT operations can be constructed. Furthermore, for this
particular class of input states it is possible to find the
general structure of all possible ideal quantum NOT operations.

\section*{Acknowledgments}
Financial support by GA\v CR 202/04/2101, by the DAAD (GA\v CR
06-01) and by the project DL 060001 of the Czech Ministry of
Education is gratefully acknowledged.

\appendix
\section{Irreducible tensor operators}
\label{tenzor} In this appendix basic properties of irreducible tensor operators
of the group SU(2) are summarized. These irreducible tensor operators are convenient
tools for implementing the covariance condition
(\ref{cavariant2}).

Rotation properties of quantum states described by the continuous
group O(3) or its universal covering group SU(2) are conveniently
analyzed by representing the density operator of this quantum
state in irreducible tensor components. In terms of orthonormal
angular momentum eigenstates $|Jm\rangle$ (with $-2J,-2J+1,...\leq
m\leq ...,2J-1,2J$ and $J$ being half integer or integer) a set of
irreducible tensor operators $T(J_{1}J_{2})_{KQ}$ (with $|J_1-J_2|
\leq K \leq J_1+J_2$ and $-K, -K+1,... \leq q \leq ...,K-1, K$) is
defined by \cite{Biedenharn,Blum}
\begin{eqnarray}
\label{irreducible} T(J_{1},J_{2})_{K q} =\sum_{m_{1}
m_{2}}(-1)^{J_{1}-m_{1}}\sqrt{2K+1} \times \left(
\begin{array}{lcr}
J_{1} & J_{2} & K\\
m_{1} & -m_{2} & -q
\end{array}
\right) |J_{1} m_{1}\rangle \otimes \langle J_{2} m_{2}|.
\end{eqnarray}
The orthogonality and completeness relations of the 3j-symbol
appearing in (\ref{irreducible}) imply the ortho-normality
relations
\begin{eqnarray}
{\rm Tr}[T(J_1,J_2)_{K q}T(J_1',J_2')^{\dagger}_{K' q'}] =
\delta_{J_1 J_1'} \delta_{J_2 J_2'} \delta_{K K'} \delta_{q q'}.
\end{eqnarray}
Thereby, ${\rm Tr}$ denotes the trace over the Hilbert space
spanned by the direct sum of the angular momentum subspaces involved.
Therefore, the irreducible tensor operators of (\ref{irreducible})
may be viewed as special examples of complete orthogonal sets of
operators which have particularly simple transformation properties
with respect to  the rotation group. These transformation
properties are described by the relation
\begin{eqnarray}
U T(J_1 J_2)_{KQ}U^{\dagger} &=& \sum_{q} T(J_1 J_2)_{Kq}
D(U)_{qQ}^{(K)},\nonumber\\&& \label{irred2}
\end{eqnarray}
with $D(U)_{qQ}^{(K)}$ denoting rotation matrix elements \cite{Blum}. These latter matrix elements
fulfill the orthogonality relation
\begin{equation}
\label{rotation_orthogonality} \int
D(\gamma\beta\alpha)^{(j)*}_{mm'}
D(\gamma\beta\alpha)^{(J)}_{MM'}\sin{\beta}\hspace{0.5em} d\beta
d\alpha d \gamma =
\frac{8\pi^2}{2J+1}\delta_{jJ}\delta_{mM}\delta_{m'M'}.
\end{equation}
Thereby, $\alpha$, $\beta$, and $\gamma$ denote the Euler angles
characterizing a particular rotation. According to (\ref{irred2})
the quantum numbers $J_1$, $J_2$, and $K$ characterize a
particular irreducible representation of the rotation group.

As the tensor operators of (\ref{irreducible}) form a complete set
any operator including the density operator $\rho$ can be
decomposed according to
\begin{equation}
\rho= \sum_{J_1 J_2 K q} {\rm Tr} \left\{T(J_1 J_2)^{\dagger}_{Kq}
\hspace{0.5em}\rho \right\} T(J_1 J_2)_{Kq}.
\end{equation}
In the special case of two qubits with angular momenta
$J=\frac{1}{2}$, for example, in such a decomposition the
irreducible tensor operators $T(\frac{1}{2},\frac{1}{2})_{Kq}$ (with
$K \in \{0,1\}$ and $-K \leq q \leq K$) appear for each qubit. Their
explicit form is given by ({\ref{tensor_components}). Obviously, the
set of tensor products of irreducible tensor operators is also a
complete set of operators on the two-qubit Hilbert space and we can
express an arbitrary two-qubit density operator in the form of
(\ref{rozklad}). With the help of the relation
(\ref{rotation_orthogonality}), finally,  it is straightforward to
prove that the most general form of an output state fulfilling the
covariance condition (\ref{cavariant2}) is given by
(\ref{general_output}).

\section{Complete positivity}
\label{Appendix_a2}

In this appendix the basic steps imposed on covariant two-qubit
quantum processes by complete positivity are discussed. This can
be done in a convenient way with the help of the theorem of
Jamiolkovski and Choi \cite{Jamiolkovski,Jamiolkovski1,Choi} whose
contents is summarized in the following.

Let $H$ be an $n$-dimensional Hilbert space
with an inner product, say $\langle .|.\rangle$,
and let $\mathcal{B}(H)$
be the associated $n^2$-dimensional Hilbert space of linear operators on
$\mathcal{H}$ whose inner product $(.,.)$ is defined by the
relation $(A,B) = Tr(A^{\dagger}B)$ for all $A,B \in
\mathcal{B}(H)$. Furthermore, let $\mathcal{L}(H_{1},H_{2})$ be the vector
space of linear transformations from a $n_1$-dimensional Hilbert
space $H_1$ to a $n_2$-dimensional Hilbert space $H_2$ and let
$I\in \mathcal{L}(\mathcal{B}(H),\mathcal{B}(H))$ denote the linear
identity operation acting on $\mathcal{B}(H)$.
A
linear transformation $T \in
\mathcal{L}(\mathcal{B}(H_1),\mathcal{B}(H_2))$ is called completely
positive if the tensor product $T\otimes I$ maps an arbitrary
positive operator $A \in \mathcal{B}(H_1 \otimes H)$
 onto a positive operator $B \in \mathcal{B}(H_2 \otimes H)$ \cite{Nielsen}.

The problem to answer the question whether a given linear
operation is completely positive or not can be solved with the
help of a theorem due to Jamiolkovski and Choi
\cite{Jamiolkovski,Jamiolkovski1,Choi}. This theorem states the
following: {\theorem[Choi,Jamiolkovski]\label{theorem_choi} Let
 $\{|u_{i}\rangle\}$ 
be an arbitrary orthonormal basis in the Hilbert space $H_1$ and
$P_{ij}=|u_i\rangle \langle u_j|$ 
be the corresponding standard orthonormal basis in the Hilbert space $\mathcal{B}(H_1)$.
Then a linear operation $T \in
 \mathcal{L}(\mathcal{B}(H_1),\mathcal{B}(H_2))$ is completely
positive if and only if the linear operator $\mathcal{J}(T)=
\sum_{ij}T(P_{ij}) \otimes P_{ij}$ is positive.}

With the help of this theorem we can determine for which
parameters $(V,X,Y)$ the covariant quantum process $\Pi_{V,X,Y}$
is completely positive. The covariance condition
(\ref{cavariant2}) associates an arbitrary input state
(\ref{vstup2}) to the output state (\ref{output2}). We can express
this relation between the input and output state by the linear
transformation
\begin{equation}
\label{help2} \rho_{out}=
\Pi_{V,X,Y}\left(\rho_{in}(\vec{P},\vec{Q},\mathcal{M})\right)=
\sum_{i,j=0}^{3} l_{ij} L_{ij} \rho_{in}
(\vec{P},\vec{Q},\mathcal{M}) L_{ij}^{\dagger}
\end{equation}
with
\begin{eqnarray}
l_{00}&=&\frac{1}{16}(1+3X+3V+9Y), \hspace{2em}
l_{i0}=\frac{1}{16}(1+3X-V-3Y), \nonumber \\
l_{0i}&=&\frac{1}{16}(1+3V-X-3Y), \hspace{2.5em}
l_{ij}=\frac{1}{16}(1-X-V+Y),
\end{eqnarray}
and with
\begin{eqnarray}
L_{00}&=& I \otimes I, \hspace{2em} L_{i0}=\sigma_i \otimes I,
\nonumber \\
L_{0i}&=&I \otimes \sigma_i, \hspace{1.5em} L_{ij}=\sigma_i
\otimes \sigma_j.
\end{eqnarray}
If $l_{ij} \geq 0$ for all $i,j\in \{ x,y,z\}$ the covariant
process $\Pi_{V,X,Y}$ is completely positive and the Kraus
operators can be written in the form (\ref{krausform}). Therefore
the conditions (\ref{relace_positivity}) are sufficient to
guarantee the complete positivity of the operator $\Pi_{V,X,Y}$.
That these conditions are also necessary follows from theorem
\ref{theorem_choi}. With the aid of (\ref{help2}) one can check
easily that the eigenvalue spectrum of the operator
$\mathcal{J}(\Pi_{V,X,Y})= \sum_{ij=1}^4\Pi_{V,X,Y}(P_{ij})
\otimes P_{ij}$ is given by
\begin{equation}
\sigma\left( \mathcal{J}(\Pi_{V,X,Y}) \right)=
\left\{\frac{1}{4}(1+3X+3V+9Y),\frac{1}{4}(1+3X-V-3Y),\frac{1}{4}(1+3V-X-3Y),\frac{1}{4}(1-X-V+Y)
\right\}.
\end{equation}
Hence, the covariant process $\Pi_{V,X,Y}$ is completely positive
if and only if the conditions (\ref{relace_positivity}) are
fulfilled.

\section{The error measure and its covariant optimality}
In this appendix the relation (\ref{distance2}) is proved for the
error measure and it is shown that this error measure does not
depend on the pure two-qubit input state selected but only on its
degree of entanglement. Furthermore, for the sake of completeness
we recapitulate the proof that whenever there is an optimal
quantum NOT operation at all, then there exists also an associated
covariant one.

\subsection{Basic properties of the error measure}
\label{appendix2a}

Let us first of all prove equation (\ref{distance2}). We start
from an arbitrary two-qubit density operator $\rho$. Let us denote
the eigenvectors of its restriction onto the three dimensional
subspace orthogonal to $|\phi\rangle$,
$\mathcal{H_{\phi}}^{\bot}$, by $|\phi_1\rangle$,
$|\phi_2\rangle$, and $|\phi_3\rangle$. The orthonormal vectors
$|\phi\rangle$, $|\phi_1\rangle$, $|\phi_2\rangle$, and
$|\phi_3\rangle$ form an orthonormal basis in which this density
operator takes the form
\begin{equation}
\rho= \left(  \begin{array}{cccc} \lambda_1 & \lambda_2 &
\lambda_3
& \lambda_4 \\
\lambda_2^{*} & \beta_1 & 0 & 0\\
\lambda_3^{*} & 0 & \beta_2 & 0 \\
\lambda_4^{*} & 0 & 0 & \beta_3 \end{array} \right), \hspace{1em}
{\rm with} \hspace{2em} \lambda_1 + \sum_{i=1}^3\beta_i =1
\hspace{2em} \lambda_1,\beta_i \geq 0.
\end{equation}
The coefficients $\lambda_i$ and $\beta_i$ are restricted by the
requirement of positivity of $\rho$. In this base an arbitrary
quantum state which is located entirely in the orthogonal subspace
spanned by the states $|\phi_1\rangle$, $|\phi_2\rangle$, and
$|\phi_3\rangle$ can be represented by a matrix of the form
\begin{equation}
\sigma= \left(  \begin{array}{cccc} 0 & 0 & 0
& 0 \\
0 & \alpha_{11} & \alpha_{12} & \alpha_{13}\\
0 & \alpha_{12}^{*} & \alpha_{22} & \alpha_{23} \\
0 & \alpha_{13}^{*} & \alpha_{23}^{*} & \alpha_{33} \end{array}
\right), \hspace{1em} {\rm with} \hspace{2em}
\sum_{i=1}^3\alpha_{ii} =1, \hspace{2em} \alpha_{ii} \geq 0.
\end{equation}
Again the coefficients $\alpha_{ij}$ have to be consistent with the positivity of $\sigma$.
In this notation
the measure $D(\rho|\phi^{\bot})$ assumes the form
\begin{eqnarray}
\label{diag} D(\rho|\phi^{\bot})&=&\min_{\sigma \in
\Gamma(\mathcal{H_{\phi}}^{\bot})} Tr\{\rho -
\sigma\}^2=\min_{\sigma \in \Gamma(\mathcal{H_{\phi}}^{\bot})}
\left\{Tr(\rho^2) - 2 Tr(\rho\sigma) + Tr(\sigma^2) \right\}
\nonumber \\
&=& \min_{\sigma \in \Gamma(\mathcal{H_{\phi}}^{\bot})} \left\{
Tr(\rho^2) -2\sum_{i=1}^3\beta_i \alpha_{ii} + \sum_{i=1}^3
\alpha_{ii}^2 + 2\sum_{i,j=1; i < j}^3 |\alpha_{ij}|^2 \right\}
\nonumber \\
 &=&\min_{\sigma \in
{\rm diag}\Gamma(\mathcal{H_{\phi}}^{\bot})} \left\{ Tr(\rho^2)
-2\sum_{i=1}^3\beta_i \alpha_{ii} + \sum_{i=1}^3 \alpha_{ii}^2
\right\}.
\end{eqnarray}
In the last equation we used the fact that the minimum is achieved
on the set of density matrices $\Gamma(\mathcal{H_{\phi}}^{\bot})$
which are diagonal in the base $|\phi\rangle$, $|\phi_1\rangle$,
$|\phi_2\rangle$, $|\phi_3\rangle$. The set of these density
operators we denoted by ${\rm
diag}\Gamma(\mathcal{H_{\phi}}^{\bot})$. Therefore, the quantity
(\ref{diag}) has to be minimized with respect to nonnegative
coefficients $\alpha_{ii}$ constrained by the condition
$\sum_{i=1}^3 \alpha_{ii} =1$. Using the method of Lagrangian
multipliers one obtains the minimum at the point $\alpha_{ii} =
\beta_i + \frac{1}{3}\lambda_1$ and its value is given by
\begin{equation}
\label{distance3}
 D(\rho|\phi^{\bot})=2\sum_{i=2}^4|\lambda_i|^2+\frac{4}{3}\lambda_1^2 =
 2\sum_{i=2}^4|\langle \phi|\rho|\phi_i \rangle|^2
 +\frac{4}{3}\langle \phi|\rho|\phi\rangle^2.
 \end{equation}
This expression can also be rewritten in the equivalent form
\begin{eqnarray}
D(\rho|\phi^{\bot})=2\left\{\sum_{i=2}^4 \langle
\phi|\rho|\phi_i\rangle\langle \phi_i|\rho|\phi\rangle + \langle
\phi|\rho|\phi\rangle\langle
\phi|\rho|\phi\rangle\right\}-\frac{2}{3}\langle
\phi|\rho|\phi\rangle^2 = 2\langle
\phi|\rho^2|\phi\rangle-\frac{2}{3}\langle
\phi|\rho|\phi\rangle^2.
\end{eqnarray}
This form (compare with (\ref{distance2})) explicitly exhibits the
independence of this measure on the diagonalization procedure used
in its derivation.

From equation (\ref{distance2}) it is straightforward to prove
that the distance $D(\rho|\phi^{\bot})$ for covariant processes
(\ref{krausform}) is unbiased with respect to all states from a
given class $\Omega_{\alpha}$. Suppose we have an arbitrary
covariant process $\Pi$ and an input state $|\phi \rangle \in
\Omega_{\alpha}$. We denote its associated output state by
$\rho_{\phi}$ ($\rho_{\phi}=\Pi(|\phi\rangle \langle \phi|)$). Let
us now take another input state $|\psi \rangle \in
\Omega_{\alpha}$ connected with the state $|\phi\rangle$ by a
unitary transformation $U=U_1\otimes U_2$ ($U_1,U_2 \in SU(2)$).
The distance $D(\rho_{\psi}, \psi^{\bot})$ between this state and
its associated output state $\rho_{\psi} = \Pi(|\psi\rangle
\langle \psi|)$ is given by
\begin{eqnarray}
\label{distance_covariant}
D\left(\rho_{\psi},\psi^{\bot}\right)&=& 2\langle
\psi|\rho_{\psi}^2|\psi\rangle-\frac{2}{3}\langle
\psi|\rho_{\psi}|\psi\rangle^2 = 2\langle \psi|\Pi(|\psi\rangle
\langle \psi|)^2|\psi\rangle-\frac{2}{3}\langle
\psi|\Pi(|\psi\rangle \langle \psi|)|\psi\rangle^2 \nonumber \\
&=& 2\langle \phi|U^{\dagger} \Pi(U|\phi \rangle \langle
\phi|U^{\dagger} )^2U|\phi\rangle-\frac{2}{3}\langle
\phi|U^{\dagger}\Pi(U|\phi\rangle \langle
\phi|U^{\dagger})U|\phi\rangle^2.
\end{eqnarray}
With the help of the covariance condition (\ref{cavariant2}) this
expression can be rewritten in the form
\begin{eqnarray}
D\left(\rho_{\psi},\psi^{\bot}\right) = 2\langle \phi|U^{\dagger}
U\Pi(|\phi \rangle \langle \phi|)^2U^{\dagger}
U|\phi\rangle-\frac{2}{3}\langle \phi|U^{\dagger}U\Pi(|\phi\rangle
\langle \phi|)U^{\dagger}U|\phi\rangle^2 =
D\left(\rho_{\phi},\phi^{\bot}\right).
\end{eqnarray}
Hence, a covariant quantum operation yields the same error
(\ref{distance2}) for all states of a given entanglement class
$\Omega_{\alpha}$.

\subsection{Optimality of covariant maps}
\label{appendix2b} Let us prove the statement that an optimal
quantum NOT operation can always be represented by a corresponding
covariant quantum map with the same error. This proof is based on
the well-known approach used by Werner \cite{Werner} in the
context of optimal cloning of arbitrary d-dimensional quantum
states. The crucial point of this proof is the fact that for an
arbitrary and in general non-covariant quantum NOT operation
$U_{\alpha}$ acting on two qubits one can define its associated
average $\widehat{U_{\alpha}}$ over all group operations
\begin{equation}
\widehat{U_{\alpha}}(\rho) = \int dU_1dU_2 \hspace{0.5em}
(U_1^{\dagger}\otimes U_2^{\dagger}) U_{\alpha}\left(U_1\otimes
U_2 \rho U_1^{\dagger}\otimes U_2^{\dagger}\right)(U_1\otimes
U_2),
\end{equation}
where $dU_1dU_2$ denotes the normalized left invariant Haar
measure of the group $SU(2)\otimes SU(2)$. The resulting quantum operation is also an
admissible NOT operation and, in addition, it also
fulfills the covariance condition (\ref{cavariant2}).
For a quantum NOT operation $U_{\alpha}$ our error measure reads
\begin{equation}
\Delta(U_{\alpha})=\sup_{|\phi\rangle \in
\Omega_{\alpha}}D(\rho_{out}|\phi^{\bot})=\sup_{|\phi\rangle \in
\Omega_{\alpha}} \left\{2\langle
\phi|\rho_{out}^2|\phi\rangle-\frac{2}{3}\langle
\phi|\rho_{out}|\phi\rangle^2\right\}.
\end{equation}
This error is a convex function of the quantum operation $U_{\alpha}$. This can be seen by
considering
a convex combination of two arbitrary two-qubit quantum
operations, say $V_1$ and $V_2$, and an
arbitrary two-qubit pure input state, say $\sigma = |\phi\rangle\langle
\phi|$. The distance $D(\rho_{out}|\phi^{\bot})$
fulfills the inequality
\begin{eqnarray}
D(\eta V_1(\sigma) +(1-\eta)V_2(\sigma)|\phi^{\bot})&=&\eta
D(V_1(\sigma)|\phi^{\bot})+(1-\eta)D(V_2(\sigma)|\phi^{\bot})-\eta(1-\eta)D(V_1(\sigma)+V_2(\sigma)|\phi^{\bot})
\nonumber \\ &\leq& \eta
D(V_1(\sigma)|\phi^{\bot})+(1-\eta)D(V_2(\sigma)|\phi^{\bot})
\end{eqnarray}
and is therefore  convex. Our error measure $\Delta$ is defined as
the supremum of a set of convex expressions in $U_{\alpha}$ and
hence is also convex. This implies the inequality
\begin{equation}
\label{cov_optimality}
\Delta(\widehat{U_{\alpha}}) \leq
\Delta(U_{\alpha}).
\end{equation}
Therefore, optimal quantum NOT operations which minimize the error
can always be found in the form of covariant quantum processes
fulfilling (\ref{cavariant2}).

\section{Determination of the optimal two-qubit quantum NOT operation}
\label{appendix3}

In this appendix the optimal two-qubit quantum NOT operations are
determined for all values of the entanglement parameter $0\leq
\alpha\leq 1/\sqrt{2}$. For this purpose we have to minimize the
error of (\ref{fidelity_unot}) under the constraints of complete positivity as
given by the relations (\ref{relace_positivity}).

Let us first of all consider the case of
non-entangled states, i.e. $\alpha =0$. The lower bound of the error
(\ref{fidelity_unot}) can be derived with the help of
inequality (\ref{relace_positivity}), i.e. $Y \geq -\frac{1}{9}
-\frac{1}{3}(X+V)$, which yields
\begin{equation}
\label{odhad_F1}
 \Delta(Z=V+X,Y) \geq \frac{4}{3} \left\{\frac{1}{6}
(Z)(1-10\alpha^2\beta^2)+\frac{2}{9}(1-\alpha^2\beta^2)\right\}^2.
\end{equation}
Minimizing the right hand side of inequality (\ref{odhad_F1})
with respect to the parameters $X$ and $V$ yields the minimal error
\begin{equation}
\Delta_{0} = \frac{4}{243}
\end{equation}
for $X=V=-\frac{1}{3}$. Hence, from relations
(\ref{relace_positivity}) we obtain the result $Y=\frac{1}{9}$.

The same approach can be used for maximally entangled states with
$\alpha =1/\sqrt{2}$. Now, an estimation of a lower bound can be
based on inequality (\ref{relace_positivity}) rewritten in the
form $X+V \geq -\frac{1}{3} - 3Y$. The resulting lower bound is
given by
\begin{equation}
\label{odhad_F2}
 \Delta(Z=V+X,Y) \geq
\frac{4}{3} \left\{
\frac{1}{6}(1+2\alpha^2\beta^2)+\frac{1}{2}Y(-1+10\alpha^2\beta^2)\right\}^2.
\end{equation}
The minimization of this lower bound leads to the
minimal error
\begin{equation}
\Delta_{1/\sqrt{2}} = 0.
\end{equation}
It is achieved for quantum processes characterized by
parameters $(V,X,Y)$ which are element of the line segment
$Y=-\frac{1}{3}$, $X+V=Z = -\frac{1}{3}$, and $X+V=\frac{2}{3}$.

Let us now consider the general case $\alpha \in (0,1/\sqrt{2})$.
Local extrema of relation (\ref{fidelity_unot}) are determined by
the conditions
\begin{equation}
\frac{\partial \Delta(Z=V+X,Y)}{\partial Z} =0 \hspace{1em} \wedge
\hspace{1em} \frac{\partial \Delta(Z=V+X,Y)}{\partial Y} =0
\hspace{1em} \Rightarrow \hspace{1em} V=X=Y=-\frac{1}{3}.
\end{equation}
The point $V=X=Y=-1/3$ at which this local minimum is reached is not
contained in the tetrahedron $\overline{ABCD}$. Therefore, the
minimum error has to be attained at points of the triangles which
form the surface of the tetrahedron $\overline{ABCD}$. It can be
checked in a straightforward way that the minima for all values of
$\alpha \in (0,1/\sqrt{2})$ are contained in the triangle
$\overline{ABC}$. This latter triangle is defined by the relation
$Z=X+V=-3Y-\frac{1}{3}$ with $-\frac{1}{3} \leq Y \leq \frac{1}{9}$
and $-\frac{1}{3} \leq X,V \leq 1$. With the help of the
substitution $Z = -3Y-\frac{1}{3}$ in (\ref{fidelity_unot}) we
obtain a quadratic function of $Y$ which is minimal at the point
\begin{equation}
Y_{min}=-\frac{1}{3}  \hspace{0.5em}
\frac{2-31\alpha^2\beta^2+20\alpha^4\beta^4}{-2-35\alpha^2\beta^2+100\alpha^4\beta^4}.
\end{equation}
This condition is valid for all values of $\alpha \in
(0,1/\sqrt{2})$. However, the relation $Y \leq 1/9$ is valid only
as long as $\alpha \geq \alpha_0$ with
$\alpha_0=\sqrt{(1-\sqrt{1-4K})/2}$ and $K=(8-3\sqrt{6})/20$. The
minimal error in the range $\alpha \leq \alpha_0$ is achieved by
the largest $Y$ value satisfying the condition $Y \leq 1/9$, i.e.
by $Y= 1/9$. As a result we obtain the relation
\begin{equation}
\label{appen_error}
 \Delta_{\alpha} = \left\{ \begin{array}{ccc}
\frac{1}{243}
\left(4+160\alpha^2\beta^2-128\alpha^4\beta^4\right),
\hspace{1em}& Y=\frac{1}{9}, X=V=-\frac{1}{3},
X+V=-3Y-\frac{1}{3},\hspace{1em} & {\rm for} \quad \alpha \leq \alpha_0 \nonumber \\
\frac{4\alpha^2\beta^2(1-4\alpha^2\beta^2)}{2+35\alpha^2\beta^2-100\alpha^4\beta^4},
& Y_{min}, X+V=-3Y-\frac{1}{3}, & {\rm for} \quad \alpha \geq
\alpha_0
\end{array} \right..
\end{equation}
From (\ref{appen_error}) we can easily determine the value of
$\alpha$ for which $\Delta_{\alpha}$ is maximal. This happens at
$\alpha_{max} = \sqrt{1/2-\sqrt{3/20}}$. The corresponding maximum
error is given by $\Delta_{\alpha_{max}}=\frac{4}{75}$ and the
associated optimal quantum NOT operation is characterized by the
parameter range $Y=-1/15$, $X+V=-2/15$ with $-1/3 \leq X,V \leq
1$.

\end{document}